# Highly Enhanced robust room temperature ferromagnetism in CVD-grown nano-dimensional MoS$_2$ flakes by modifying edges and defect engineering

*Sharmistha Dey[a], Nahid Chaudhary[b], Ulrich Kentsch[c], Rajendra Singh[b], Pankaj Srivastava[a], and Santanu Ghosh[a*]*

[a] Nanostech Laboratory, Department of Physics, Indian Institute of Technology Delhi, New Delhi 110016, India

[b] Department of Physics, Indian Institute of Technology Delhi, New Delhi 110016, India

[c] Institute of Ion Beam Physics & Materials Research, Helmholtz-Zentrum Dresden-Rossendorf, Dresden 01314, Germany

**Abstract:**

The alterations in the magnetic properties and electronic structure of chemical vapor deposition (CVD) grown nano-dimensional molybdenum disulfide (MoS$_2$) after low energy ion irradiation are thoroughly investigated. The formation of pure hexagonal 2-H phase has been identified by Raman spectroscopy and X-ray diffraction (XRD). The pristine samples are irradiated by Argon (Ar) ions with low energy at different fluences. A comprehensive analysis of Raman spectroscopy data manifests the formation of lattice defects like S-vacancies across the samples after irradiation. Triangular-flake formation in the pristine sample is confirmed by field emission scanning electron microscopy (FESEM) images. After increasing irradiation fluences the big flakes commenced to fragment into smaller ones enhancing the number of edge-terminated structures. The electron probe microanalyzer (EPMA) analysis verifies the absence of any magnetic impurity. Rutherford backscattering spectrometry (RBS) and X-ray photoelectron spectroscopy (XPS) study confirm the formation of S-vacancies after irradiation. The pristine sample exhibits diamagnetic behavior at room temperature. The saturation magnetization value increases with increasing the ion irradiation fluences, and the sample irradiated with $1\times10^{15}$ ions/cm$^2$ demonstrates the highest magnetization value of 4.18 emu/g. The impact of edge-terminated structure and point defects like S-vacancies to induce room-temperature ferromagnetism (RTFM) is thoroughly examined.

**Corresponding author:** santanu1@physics.iitd.ac.in

# 1. Introduction:

The discovery of graphene has initiated an entirely novel era for two-dimensional (2D) materials. However, most electronic applications necessitate a band gap between the conduction and valence bands, whereas graphene has a zero-band gap [1,2]. To overcome this issue, a comprehensive investigation has been conducted into 2D transition metal dichalcogenides (TMD) resembling graphene. TMDs are typically semiconductors exhibited as $MX_2$, where M denotes the transition metals such as Mo and W, and X indicates the chalcogen like S, Se [3]. Among TMDs, molybdenum disulfide ($MoS_2$) has drawn a lot of attention due to its unique characteristics, such as photoconductive [4], effective catalytic [5], optoelectronic [6], and lubricant properties [7]. It possesses a layered structure, a thickness-dependent bandgap (1.2 eV indirect bandgap for the bulk and 1.9 eV direct bandgap for the monolayer), an intralayer strong covalent bond (between Mo and S atoms), and an interlayer weak van der Waals interaction [8]. It has strong spin splitting (148 meV) of the topmost valence band at the K point, which is ideal for spintronics applications [3]. The properties of nano-dimensional $MoS_2$ are distinct from the bulk due to the quantum confinement effect [9]. Bulk $MoS_2$ exhibits a diamagnetic nature. However, edge-terminated nanostructured $MoS_2$ thin films exhibit room temperature ferromagnetism (RTFM) due to their different coordination geometry and stoichiometry than bulk. In bulk $MoS_2$, spin pairing occurs due to the trigonal-prismatic coordination, but there is spin polarization in edge-terminated structures due to the octahedral coordination geometry [10].

Ferromagnetic semiconductors are essential for spintronics, as electronic charge and spin are simultaneously needed in such applications [11,12]. To convert $MoS_2$ into a magnetic semiconductor, transition metal (TM) doping-induced magnetism is extensively investigated [13–17]. Even though a substantial amount of magnetization is produced, it is impossible to determine whether the magnetism is induced by the aggregation of the TM itself due to the different crystal structures or coming intrinsically from the system [18]. Defects always exist in the solid and can change the optical, electrical, and magnetic properties due to disruption in lattice symmetry [19]. Defect-induced room temperature mmagnetism is reported in graphene, offering a novel research perspective to the researchers [20–22]. It is of great interest to investigate why nanostructured TMD semiconductors exhibit defect-induced magnetism, as 2D TMDs are similar to graphene. A comprehensive investigation of defect-induced RTFM in $MoS_2$ has been established. The reasons for the induced magnetism reported are edge states, Mo vacancies, S vacancies, S substitute Mo vacancies, Mo substitute S vacancies, generation and reconstruction of edge states, vacancy clusters, and lattice distortion, etc [23–27]. Defects can be generated by different techniques, such as changing deposition parameters and annealing temperatures [28], plasma treatment [29,30], ion irradiation [25], etc. Among these techniques, low-energy ion irradiation (LEII) is a distinctive process of altering the structural and other physical properties by creating defects. In this process, one can control the generation of defects by changing ion type, ion-irradiation energy, and fluences. In a low-energy regime, ions transfer their energy to the solid by the elastic collision process, which is referred to as nuclear energy loss ($S_n$). This process results in various point defects, such as vacancies and interstitials, as a consequence of the collision cascade [18,31]. In recent work, we have seen that the saturation magnetization value is enhanced due to the modification of edge-terminated structure and the creation of S-vacancies and S-H bond formation after low-energy hydrogen ion irradiation and hydrogen annealing in vertical edge-oriented $MoS_2$ [32].

The primary motivation of this work is to investigate the origin of enhancement in RTFM and its correlation with the electronic structure of the nano-dimensional pristine and Ar ion-irradiated $MoS_2$ films. CVD-grown nano-dimensional $MoS_2$ thin films are produced by optimizing all the growth parameters in a single-zone, mini-programmable tube furnace. The pristine sample behaves as a diamagnetic sample as it does not contain a sufficient number of edge-terminated structures and S-vacancies. The pristine samples are irradiated with 10 keV $Ar^+$ ions with a tilt angle of 7° with different fluences, such as $1 \times 10^{13}$, $5 \times 10^{13}$, $1 \times 10^{14}$, $5 \times 10^{14}$, $1 \times 10^{15}$, and $5 \times 10^{15}$ ions/cm$^2$, respectively. The $1 \times 10^{15}$ ions/cm$^2$ irradiated sample shows the highest saturation magnetization value of 4.18 emu/g. One of the most crucial factors behind it is the increase in the edge-terminated structure (as seen in the FESEM images), which provides dangling bonds and free spin. Second is the increase in the number of sulfur vacancies, which is confirmed by the XPS analysis. As the films are very thin (~20 nm), the number of edges and S-vacancies per unit volume, are high, which further enhances the magnetization value. The structural modification (i.e. breaking of a triangular flake into smaller flakes), controlling the generation of edges and S-vacancies by LEII, and consequent enhancement of magnetization to a significantly higher value are rarely found in the literature.

## 2. Experimental details:

The CVD technique is employed to synthesize nano-dimensional $MoS_2$ films. The growth is carried out on a substrate of $SiO_2$ (300 nm, thermally oxidized silicon)/Si. The process utilized sulfur (S) and molybdenum trioxide ($MoO_3$) as precursors, both sourced from Sigma-Aldrich, within a single-zone mini-CVD system provided by MTI Corporation, USA. To prepare the substrate, it undergoes sequential cleaning using acetone, isopropyl alcohol (IPA), and deionized (DI) water. The amounts of the precursors used are 100 mg of sulfur and 15 mg of $MoO_3$. The growth process starts by placing $MoO_3$ powder in a crucible at the centre of the CVD tube (900 °C), which corresponds to the hottest zone. A second crucible containing sulfur powder is positioned approximately 14 cm away, in a relatively cooler region of the tube (Figure 1(a)), and the substrate is placed face down on the MoO$_3$ powder. To remove residual precursors, dust, and moisture from the system, argon (Ar) gas is initially flowed at 500 standard cubic centimeters per minute (sccm) for 10 minutes at 300 °C. For the remainder of the process, the Ar flow rate is reduced to 120 sccm. Sulfur begins melting around and fully vaporizes at approximately 600 °C and 900 °C (the temperature at the central zone). The growth of MoS$_2$ occurred at a temperature of 900 °C for 10 minutes. Thereafter, the system is left to cool naturally.

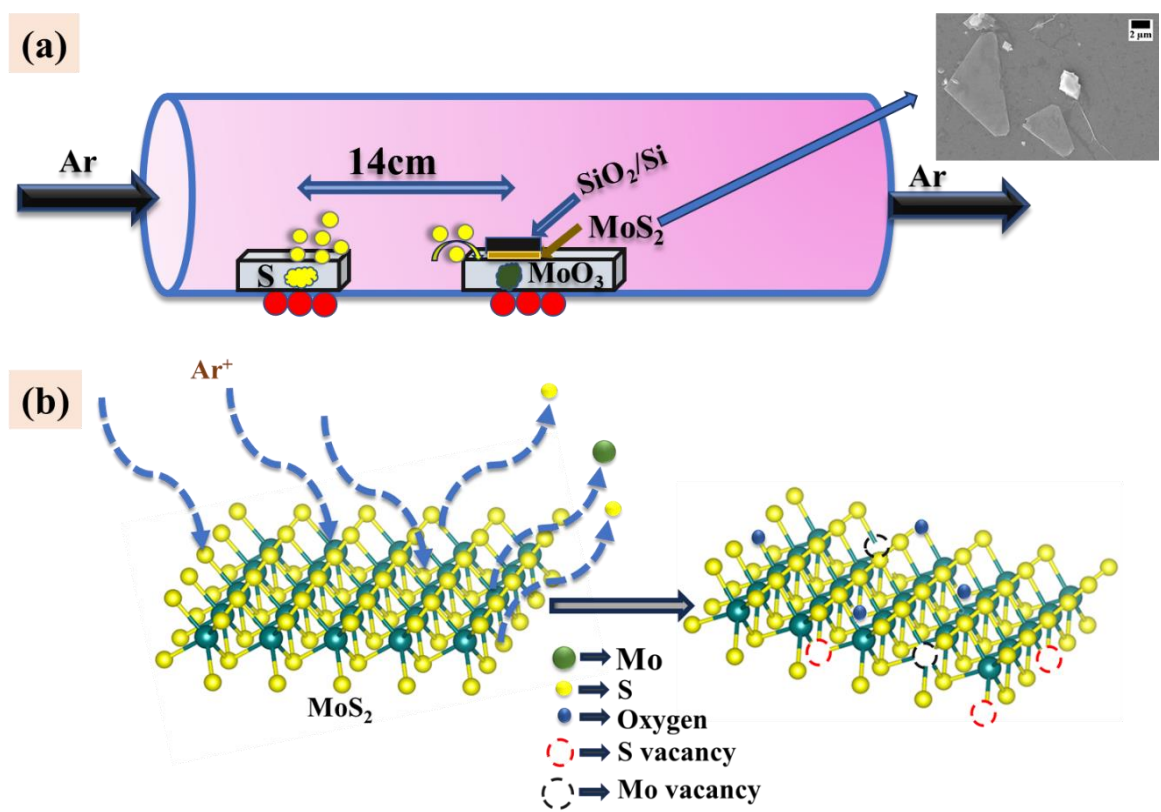

**Figure 1.** (a) A schematic illustration of the CVD system used for depositing nanostructured MoS$_2$ thin films. (b) A schematic representation of Ar$^+$ irradiation on MoS$_2$, leading to the formation of S and Mo vacancies.

Subsequently, 10 keV inert Ar ions are used to irradiate CVD-grown triangular flakes of MoS$_2$ with a tilt angle of 7° with different irradiation doses. The projected ion ranges (R), and the nuclear and electronic energy loss (S$_n$ and S$_e$) are calculated by the Stopping and Range of Ions in Matter (SRIM) simulation software [33–35]. For 10 keV Ar$^+$ ion irradiation with a tilt angle of 7° in MoS$_2$ (The density of MoS$_2$ is taken as 5.06 g-cm$^{-3}$), the projected ranges with straggling, S$_n$, and S$_e$ are 10 $\pm$ 5 nm, 0.69, and 0.14 keV/nm respectively (Figure S1). Hereafter, the pristine, 10 keV Ar$^+$ irradiated samples with irradiation fluences $1 \times 10^{13}$, $5 \times 10^{13}$, $1 \times 10^{14}$, $5 \times 10^{14}$, $1 \times 10^{15}$, and $5 \times 10^{15}$ ions/cm$^2$ will be denoted by P0, P1, P2, P3, P4, P5, and P6 respectively. The detailed instrumental specification (XRD, Raman, FESEM, EPMA, MPMS3, RBS, and XPS) is discussed in the supporting information.

## 3. Results and discussion:

Figure 2 illustrates the room temperature Raman spectra of the pristine and 10 keV Ar ion irradiated MoS$_2$ samples grown on SiO$_2$/Si substrates with different fluences. The peak is located at ∼383 cm$^{-1}$ corresponds to $E_{2g}^1$ peak, produced due to the in-plane vibration of Mo and S atoms, and the peak situated at∼408 cm$^{-1}$ is corresponding to A$_{1g}$ peak, generating due to the out-of-plane vibration of S atoms. The A$_{1g}$ peak is more intense here, as A$_{1g}$ phonons represent atomic displacements along the c-axis and can couple more strongly with the excited $d_z^2$ states than $E_{2g}^1$ phonons [36,37]. The 2-H MoS$_2$ phase formation is confirmed from the Raman spectra, as no other impurity peak is observed. The peak at ∼450 cm$^{-1}$ corresponds to the second-order longitudinal acoustic mode (2LA), produced by the double resonance Raman process [38–40]. There are no considerable changes in intensity in the 2LA mode after irradiation, as it is not directly associated with the formation of defects [41,42]. From Figure 2(a), it is evident that a new peak at ∼220 cm$^{-1}$ emerged, which corresponds to the first-order LA mode. This peak is generated due to the presence of defects like S-vacancies [31,42]. The LA mode is absent in the P0 sample, but after irradiation, this peak is generated, and the intensity increases from the P1 to P5 samples. However, at higher irradiation fluences, the intensity of the LA peak decreases for the P6 sample, as the sample deteriorates.

It has been observed from Figure 2(b) that there is a left shift for both $E_{2g}^1$ and A$_{1g}$ modes. The A$_{1g}$ peak shifted from 408 cm$^{-1}$ to 403 cm$^{-1}$ and $E_{2g}^1$ peak shifted from 383 cm$^{-1}$ to 376 cm$^{-1}$ from P0 to P5 samples. It also confirms the presence of the creation of vacancies. The $E_{2g}^1$ peak is more sensitive to the creation of vacancies and other types of defects than the peak A$_{1g}$; due to the presence of S-vacancies, the length of the covalent bond between two nearby Mo atoms decreases. As a result, the in-plane vibration energy decreases more significantly than the out-of-plane vibration [42,43]. It is also observed that the line width (FWHM) of both peaks ( $E_{2g}^1$ and A$_{1g}$) is widening with increasing irradiation doses from P0 to P5 samples. There is the merging of defect peaks (satellite peaks) with the main vibrational peaks. The satellite peaks are generated due to the local phonon modes formed from the local vibration surrounding the vacancies. To see the merging of these two peaks, the Raman spectra of P0, P3, and P5 samples are deconvoluted by Lorentzian peak fitting as seen in Figures 2(d), (e), and (f). The Raman spectra of the pristine sample are fitted by only three peaks mainly the $E_{2g}^1$, A$_{1g}$, and 2LA modes. But for P3 and P5 samples, there is the generation of two new peaks $E_{2g}^{1\prime}$ and $A_{1g}^\prime$ (satellite peaks). These two peaks correspond to S and Mo vacancies. The FWHM for the $E_{2g}^1$ and the A$_{1g}$ peaks increases and intensity decreases for the P3 and P5 samples compared to the P0 sample, manifesting that there is an induced tensile strain (stretching in bond length) due to the generation of different types of vacancies (like S-vacancies) and defects [44,45]. By keeping the same FWHM for the $E_{2g}^{1'}$ and $A_{1g}'$ peaks for the P3 and P5 samples, the intensity increases for both the peaks from the P3 to P5 sample, confirming the formation of S and Mo vacancies [42].

The XRD pattern of the pristine and irradiated samples is shown in Figure S2. The peak at 14.45° corresponds to the (002) plane of MoS$_2$. It also illustrated the pure hexagonal 2-H phase formation of few-layer MoS$_2$ thin films with space group P6$_3$/mmc [46–49]. No other phases or impurity phases are observed in the XRD pattern for the pristine and irradiated samples. The intensity of the peak is comparatively low as the film contains ultrathin layers [50,51]. From Figure S2, it is evident that after Ar$^+$ irradiation there is a shift of the (002) peak in a lower Bragg angle from 14.45° to 14.12° for the P0 to P5 samples. The FWHM also gradually increases from 0.52° to 0.98° after irradiation from the pristine to the P5 sample. An increase in FWHM after irradiation manifests deterioration in crystallinity. The observed redshift on irradiation indicates the development of micro-strain of the lattice [26,44]. The average crystallite size (D) is calculated from Scherrer's formula corresponding to the (002) plane of MoS$_2$. The crystallite sizes of P0, P1, P2, P3, P4, and P5 samples are found to be 15.38, 15.05, 12.57, 9.51, 8.41, and 8.09 nm respectively. The lowering of crystallite sizes after irradiation suggests deterioration and damage in the lattice [52,53]. The values of the (002) peak position, FWHM, and crystallite size for all the samples are tabulated in Table S2.

The surface morphology of the pristine and irradiated samples is investigated by FESEM images. Figure 3 shows the FESEM pictures of the pristine and the 10 keV Ar$^+$ irradiated samples. It exhibits triangular-shaped flakes in pristine samples and modification of the flakes after irradiation. As edge-terminated structure plays a significant role in inducing ferromagnetism in MoS$_2$, it is imperative to examine the structural modification in the films, as low energy ion irradiation modifies the structure and also creates defects such as vacancies. The thickness of the pristine sample is confirmed from the height profile by atomic force microscopy (AFM) and cross-sectional FESEM (Figure S3) images. The average thickness of the sample is 20 nm. The lateral dimensions of the triangular flakes are in the micron order (~8 $\mu$m). It is evident from Figures 3(b) to 3(k) that after irradiation the roughness of the surface increases, correspondingly increasing rough active sites [54]. The roughness gradually increases up to P4 samples with increasing irradiation fluences. For the P5 sample, the smooth surface of the flakes fragmented into many smaller ones, increasing the number of active sites and exposed edges. Less secondary nucleation is expected on the edges because of their lesser thickness [10]. An extensive defect generation study has been done using the transport of ions in matter (TRIM) simulation software for an average of 100,000 ions (the density of $MoS_2$ has been taken ∼5.06 g-cm$^{-3}$), before determining the type and energy of the ion. Argon ion (Ar$^+$) has been chosen for the irradiation, as the main aim of this study is to induce magnetism in a pure system ($MoS_2$) by creating intrinsic defects and alteration of the edges. Ion ranges and detailed collision events are shown in Figure S1. The creation of S vacancies is more probable than Mo vacancies due to the comparable mass of S atoms with Ar ion, this is called preferential sputtering. The creation of defects like dislocations, interstitials, and Mo and S vacancies occurs in the $MoS_2$ lattice due to the low energy Ar ion irradiation. The local crystal structure is weakened by the displacement of the atoms, especially at the edges of the triangular flakes. This increases the possibility that the flakes will break under the strain imposed on them by the accumulation of defects. The $MoS_2$ lattice acquires the energy from the Ar ions, which breaks bonds, particularly at the boundaries of the flakes, resulting in mechanical instability and fragmentation. The triangular flakes have inherent edge effects where the bonding is weaker compared to the inner bulk regions. Ion irradiation tends to damage these edges more intensely, and the resulting defects can propagate through the material, causing the flakes to break into smaller segments. Additionally, cracks may nucleate at these weakened edges and propagate through the entire flake, leading to fragmentation [55]. The corresponding micro-strain also increases with increasing irradiation fluences, as evident from the shifting of the peak towards lower Bragg angles [44,45]. For the highest irradiation doses (sample P6), the sample deteriorated as shown in Figure 3(l).

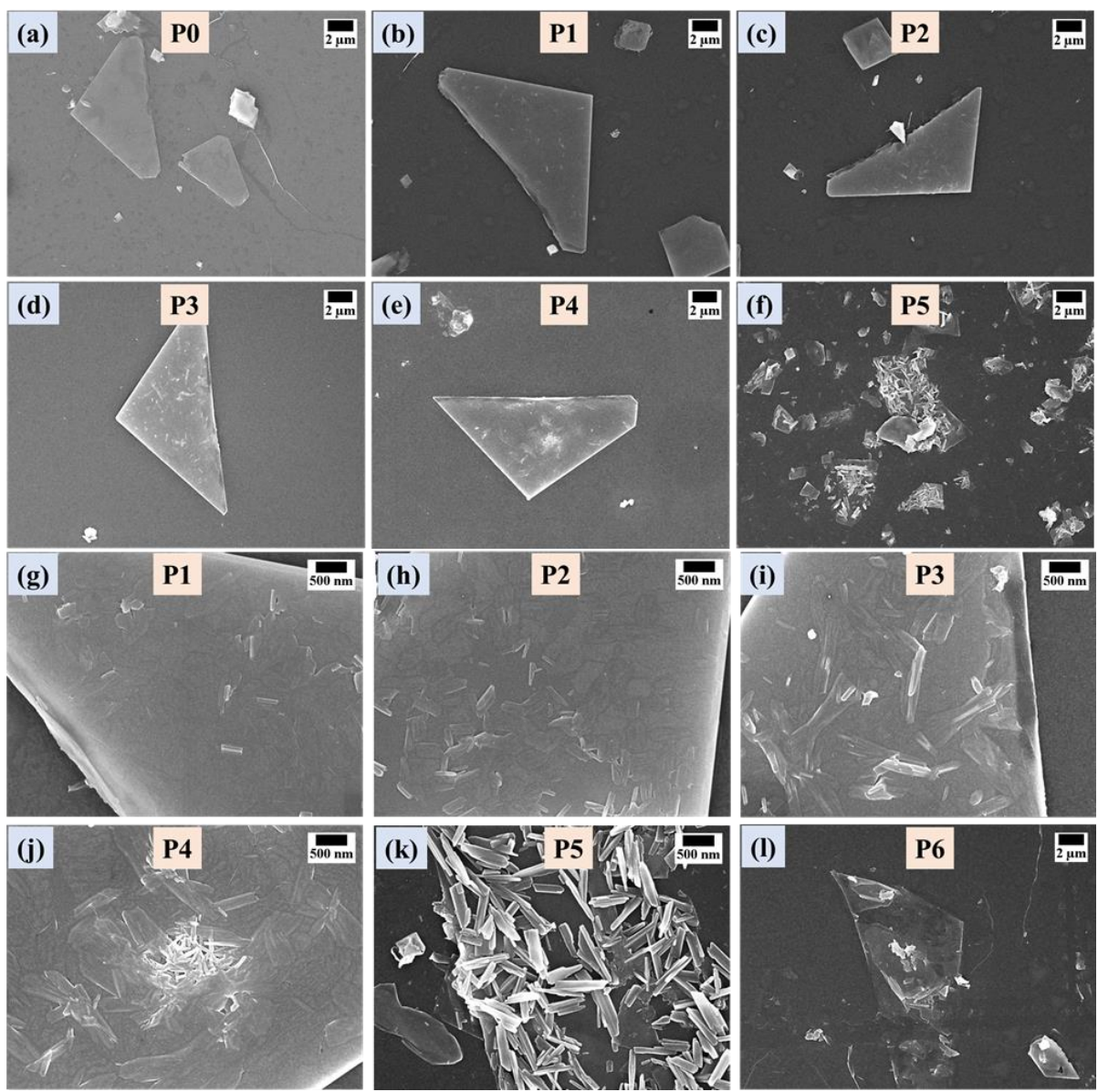

**Figure 3.** FESEM images of (a) P0, (b) P1, (c) P2, (d) P3, (e) P4, (f) P5 samples. It is evident from the figures that the surface of the sample roughened after irradiation exposing active sites. For the P5 sample, the big triangular flake fragmented into smaller ones. The zoom view of the surface is shown in (g) P1, (h) P2, (i) P3, (j) P4, (k) P5. (l) The surface of the P6 sample deteriorated by higher irradiation fluences.

The absence of any magnetic impurities in the sample is confirmed by the EPMA analysis. From the semi-quantitative analysis (Intensity versus wavelength plots), it is confirmed that only Mo and S are present in all the samples (Figure S4(d) and (e)) [56]. The Si signal is generated from the substrates, and it has no magnetic contribution. The Au signal seen here is due to coating, which is only used for FESEM and EPMA to avoid the charging effect. The distribution of Mo and S over the triangular flakes in the P0 sample is shown in Figures S4(a), (b), and (c). Initially, an area is chosen, as illustrated in Figure S4(a), and mapping is done separately for Mo and S. A PET (Pentaerythritol) crystal with the corresponding detector is used to detect Mo and S. Figure S4(b) and Figure S4(c) shows the mapping for S and Mo, respectively.

Figure 4 exhibits the magnetization (M) versus applied magnetic field (H) plots obtained from SQUID magnetometer measurements of the pristine and Ar ion irradiated samples at room temperature. The pristine sample exhibits diamagnetic behavior at room temperature. After irradiation, weak ferromagnetism is induced in the P1 sample. With increasing irradiation fluences, saturation magnetization increases up to P5 samples. The P6 sample shows again diamagnetic behavior. From Figure 4(a), it is evident that P1 to P5 samples show clear hysteresis loops with particular coercivity and remnant magnetization value; the enlarged view of the central region ($\pm$ 100 Oe) is shown in the inset. The saturation magnetization values for P1, P2, P3, P4, and P5 samples are 0.81, 2.29, 3.17, 3.64, and 4.18 emu/g, respectively. The saturation magnetization ($M_s$), remnant magnetization ($M_r$), and coercive field ($H_c$) values at 300 K are listed in Table 1. An increase of around 416 % in the Ms value is observed from P1 to P5 samples at room temperature, which is a giant enhancement in $MoS_2$. The value of Ms reported in previous literature and the present one is shown in Table S1. The low coercivity value confirms the soft ferromagnetic behavior of $MoS_2$ [57]. Although the coercivity value does not change significantly, the magnetization strength increases, which is the product of saturation magnetization and coercivity. Magnetization versus temperature measurements are shown in Figure 4(b) for the P2 sample. There is no transition temperature up to 350 K and also bifurcation in zero field-cooled (ZFC) and field-cooled (FC) curves, confirming the sample is ferromagnetic at room temperature, similar behavior is reported for other ferromagnetic samples [10,25].

The potential causes of induced magnetism in $MoS_2$ include single vacancies or vacancy clusters, edge states, and lattice reconstruction [25]. Some reports indicate that zigzag edges and defect-induced room-temperature ferromagnetism are generated in nanocrystalline $MoS_2$ films and nanoribbons [24,58]. Few theoretical [59,60] and experimental research [44,61] exist that confirm the reasons behind the induced magnetism are strain and defects. From DFT-based theoretical calculations, it is concluded that one Mo and two S vacancies give the highest magnetization value among different vacancy configurations. Also, by keeping the same vacancy concentration, the saturation magnetization value increases up to a particular thickness (12 layers) of $MoS_2$ [23]. The bulk $MoS_2$ shows diamagnetic behavior. Other studies show that edge-terminated nanostructures of $MoS_2$ show ferromagnetic behavior, and the creation of S-vacancies in this structure enhances the saturation magnetization value [26,32]. J. Zhang et al.; show from the theoretical calculations that, for the triangular edge-terminated structure, the S-edge with 100 % S coverage and the Mo-edge with 0 % S coverage show the highest magnetization value [10]. The magnetic moments reported in several research studies concerning $MoS_2$ due to defects, S-vacancies, and edge states are summarised in Table S1. The reported magnetic moment values are significantly lower than those obtained in the present work.

From the previous study, it is observed that ultrathin pristine MoS$_2$ thin films exhibit diamagnetic behavior, but after gamma-ray irradiation, they show ferromagnetism. The reason behind this is the distortion of the lattice, Mo, and S vacancies [23]. 2H-MoS$_2$ is macroscopically nonmagnetic as the Mo$^{4+}$ ions are in a trigonal prismatic local coordination, where the two 4d electrons are spin-antiparallel, resulting in a net magnetic moment of zero. If the Mo$^{4+}$ 4d electron configuration is adjusted to exhibit a nonzero magnetic moment, the 2H-MoS$_2$ phase might likewise demonstrate magnetic characteristics [62]. In the present case, edge-terminated structures and S-vacancies both contribute significantly to inducing ferromagnetism. From FESEM images, it is evident that the pristine sample contains triangular flakes. After Ar ion irradiation, the edge-terminated nanostructure increases (seen from FESEM images). For the P5 sample, one triangular flake fragmented into numerous smaller pieces, increasing the number of edge-terminated structures. Due to the preferential sputtering, S-vacancies are created in the MoS$_2$ layers. The presence of S-vacancies is confirmed by the RBS and XPS analysis, discussed in the later section. Compressive stress will be generated as a result of the excess number of vacancies. Due to this differential stress, the flakes fragmented into smaller ones [55]. The unsaturated Mo atoms at the edge states and S vacancies are the primary causes of the induced magnetism. The stoichiometry of the edge-terminated atoms differs from that of the bulk atoms. Mo atoms will be unsaturated at the edges with octahedral coordination, leading to induced ferromagnetism and spin polarization. Ferromagnetism increases with the increase in unsaturated prismatic edges. Through cooperative interaction, the spin polarization of the unpaired electrons corresponding to Mo surrounding the S-vacancy produces induced magnetic moments that display ferromagnetism. The bound magnetic polaron (BMP) model states that magnetism is induced by the spin interaction of trapped electrons at sulphur vacancies with Mo 4d ions, which results in the creation of the bound magnetic polaron.

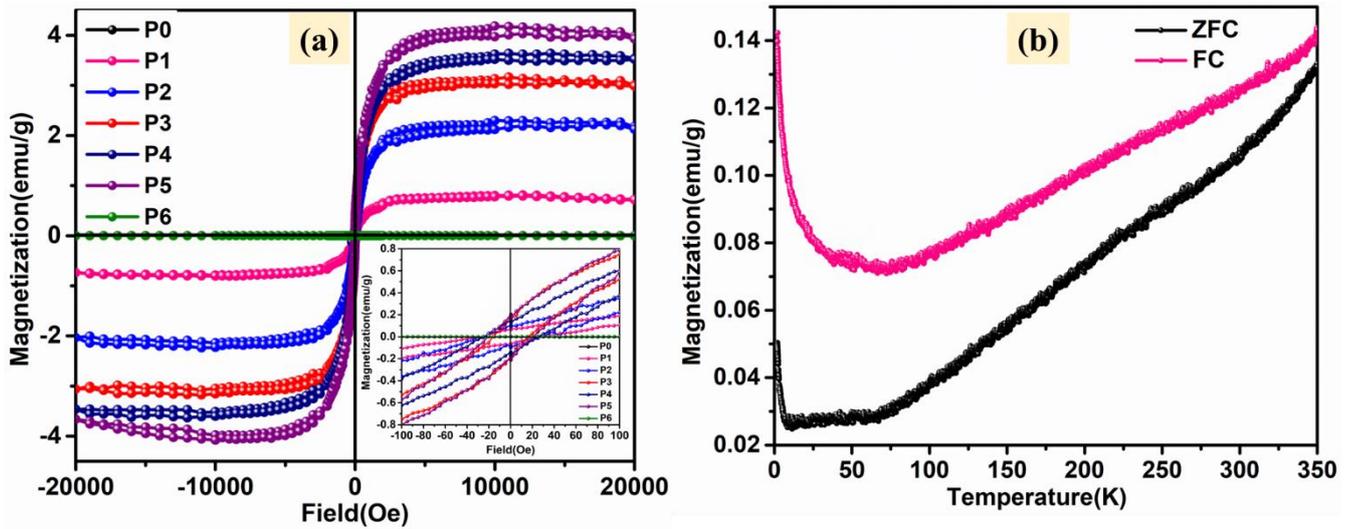

**Figure 4.** (a) M versus H plots for pristine and Ar ion irradiated samples at room temperature, the zoomed view of the central part is shown in the inset; (b) ZFC and FC plots for the P2 sample, it shows there is no transition observed up to 350 K.

Table 1. Values of Ms, Mr, and Hc of pristine, and irradiated samples at room temperature.

| Sample name | saturation magnetization ($M_s$) in emu/g | Remnant magnetization ($M_r$) in emu/g | Coercive field ($H_c$) in Oe |
|---|---|---|---|
| P0 | - | - | - |
| P1 | 0.81 | 0.071 | 25 |
| P2 | 2.29 | 0.100 | 28 |
| P3 | 3.17 | 0.150 | 24 |
| P4 | 3.64 | 0.170 | 28 |
| P5 | 4.18 | 0.220 | 25 |
| P6 | - | - | - |

Figure 5 shows the Rutherford Backscattering spectra (RBS) of the P0, P2, and P5 samples grown on $SiO_2$/Si substrate by CVD technique. There are mainly two peaks that correspond to sulfur and molybdenum elements, confirming no transition metal impurity present in the samples. The signal for oxygen originates from the substrate, while the steep edges represent the silicon substrate [63]. The experimental (black dotted line) and simulated (red line) results agreed broadly. The tailing (indicated by the red dotted circle in Figure 5) in the S and Mo peaks indicates the interdiffusion of S and Mo atoms into the substrate. The ratio of the concentration of Mo and S for the P0, P2, and P5 samples obtained from the simulated data is 1.99, 1.90, and 1.63, respectively. It confirms that after irradiation sulfur vacancies are generated in the samples. The thickness calculated from the fitted data is about 20 nm.

The electronic structure of the pristine and Ar ion-irradiated samples has been thoroughly examined by XPS to understand the increase in the magnetization value on irradiation. The Mo 3d and S 2p core-level XPS spectra of $MoS_2$ pristine and Ar ion irradiated samples are shown in a stacked representation in Figure 6. All the peaks demonstrate a shift towards lower binding energy (BE) with increasing irradiation fluences (Mo $3d_{5/2}$ shifted from 229.8 eV to 229.1 eV from P0 to P5 samples). The Mo 3d and S 2p core-level spectra also show peak broadening with increasing irradiation fluences. The shifting towards lower BE and increase in FWHM indicate the development of vacancies, defects, surface amorphization, and disorder in the irradiated samples [64,65].

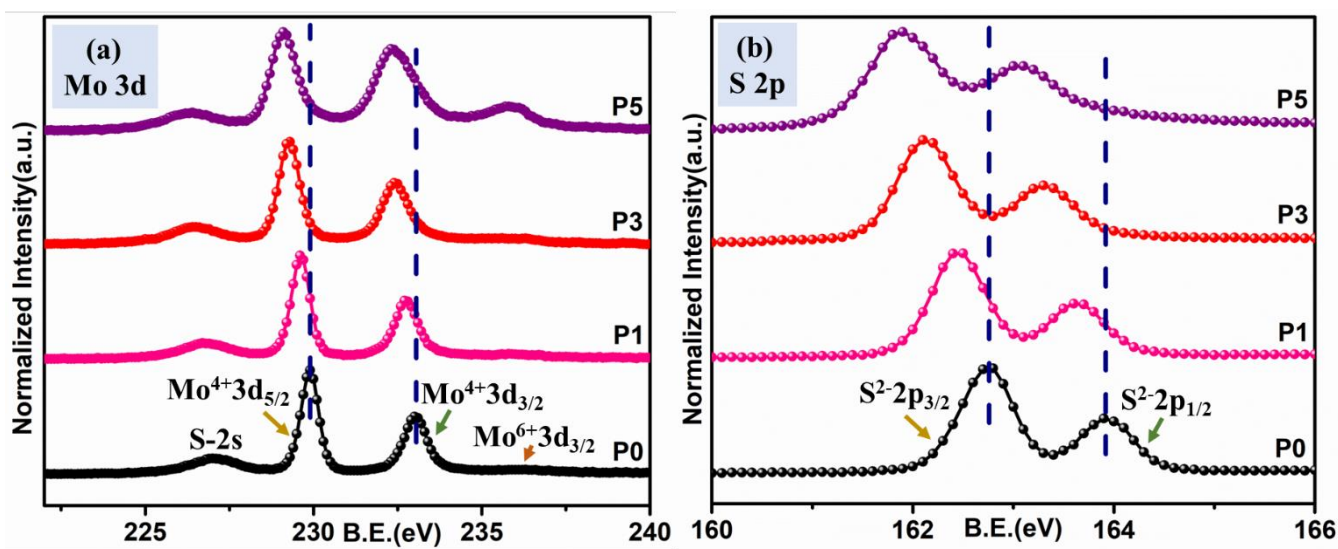

**Figure 6.** The core level XPS spectra of pristine and $Ar^+$ irradiated samples (a) Mo 3d, and (b) S 2p. The creation of S-vacancies is confirmed by the shift in the Mo 3d and S 2p spectra towards lower binding energies.

Figure 7 represents the deconvoluted Mo 3d XPS core-level spectra of the pristine and irradiated samples. The spectra are fitted into nine peaks using a pseudo-Voight function in CasaXPS software. The FWHM is maintained constant for all the corresponding peaks, and the background signals are subtracted using the iterative Tougaard method. In order to fit the Mo 3d spectrum, the distance in BE and relative area ratio between the two spin-orbit doublets ($3d_{5/2}$ and $3d_{3/2}$) have been maintained as 3.1 eV and 1.5, respectively. $Mo^{4+}$ valence states are predominant in all samples, confirming the creation of a pure $MoS_2$ phase. $Mo^{4+}$ $3d_{5/2}$ and $3d_{3/2}$ are situated at 229.9 and 233 eV, respectively; $Mo^{5+}$ $3d_{5/2}$ and $3d_{3/2}$ are positioned at 230.5 and 233.6 eV, respectively; and $Mo^{6+}$ $3d_{5/2}$ and $3d_{3/2}$ are found at 233.1 and 236.2 eV, respectively. Lower valence states of Mo, designated as a and b in Figure 7, are located at 229 eV and 232.1 eV, respectively, indicating the existence of reduced Mo, specifically the creation of S-vacancies and the development of amorphized $MoS_{2-x}$ states at the surface. The peak is located at 227 eV, indicating S 2s peak [66,67]. As ion irradiation fluences increase up to P5 samples, the peaks associated with lower valence states of Mo intensify while the peak corresponding to S 2s diminishes, signifying the emergence of S-vacancies and reduced Mo species at the surface. $MoS_2$ is a layered structured material and has van der Waals interaction between interlayers; it has a propensity to absorb oxygen at the surface [10]. It is evident that the pristine sample comprised of lesser amount of Mo 5+ and 6+ states, confirming that the samples are not oxidized appreciably at the surface and belong to the pure $MoS_2$ phase. It has been demonstrated that after ion irradiation, Mo 6+ states and lower valence states of Mo increased gradually with increasing irradiation fluences, clearly shown in Table 2. As no in situ characterization is used, the surface of the samples is exposed to environmental oxygen. It is also established that S-vacancies are increasing with increasing irradiation fluences, and further oxygen got incorporated in the S-vacancies; consequently, Mo 6+ states are increasing with increasing irradiation fluences. As XPS is a surface-sensitive technique, it can be inferred that the oxygen is incorporated at the surface only. From combined Raman, RBS, and XPS studies, it is determined that the S-vacancies are created across the entire sample, but oxygen is incorporated at the surface only.

Figure S5 displays the S 2p spectra of pristine and Ar ion irradiated samples (P3 and P5). The $2p_{3/2}$ and $2p_{1/2}$ peaks corresponding to the S 2p core level are located at 161.8 eV and 163 eV, with the distance between the BE being 1.2 eV, respectively. At higher binding energy, there is broadening after irradiation, coming from polysulfide ions ($S_2^{2-}$ or $(S-S)^{2-}$). At the time of irradiation and annealing, the sputtered or removed S atoms may form a bond with each other and form $(S-S)^{2-}$. If one Mo is sputtered out, the surrounding S atoms become unsaturated and can form a bond again, forming the $(S-S)^{2-}$. The broadening at lower binding energies indicates the presence of lower valence states of S, suggesting the formation of S-vacancies and the formation of the $MoS_{2-x}$, x>0, amorphous phase at the surface [64,65].

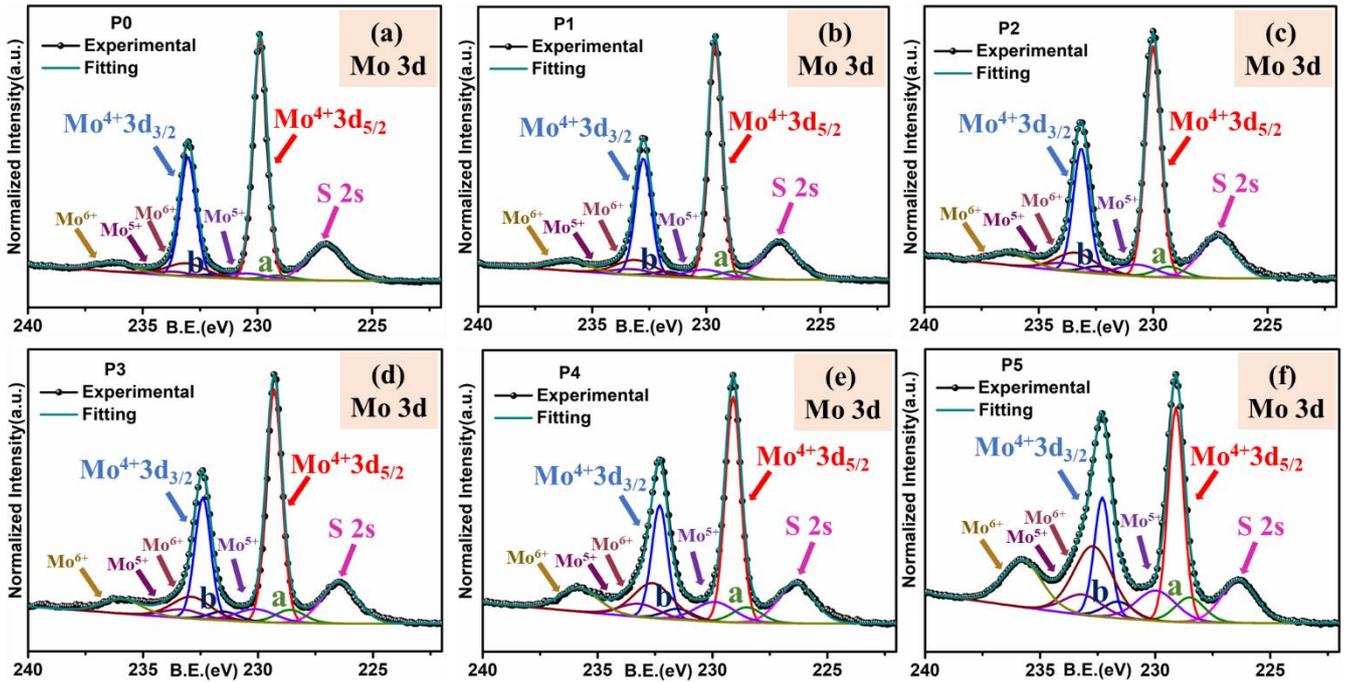

**Figure 7.** Mo 3d core level XPS spectra of (a) Pristine and (b)-(f) Ar ion irradiated samples. The intensities of peaks corresponding to lower valence states increase and the area under Mo 6+ also increases with increasing irradiation fluences.

Table 2. Mo content in Different valence states and S 2s content for pristine and irradiated samples.

| Samples | $Mo^{4+}$ content (%) | $Mo^{5+}$ content (%) | $Mo^{6+}$ content (%) | Lower valance state of Mo content (%) | S 2s content (%) |
|---|---|---|---|---|---|
| P0 | 65.88 | 4.62 | 10.38 | 2.77 | 16.35 |
| P1 | 61.50 | 6.58 | 11.58 | 4.40 | 15.95 |
| P2 | 58.29 | 7.85 | 12.91 | 5.48 | 15.47 |
| P3 | 56.24 | 8.33 | 14.23 | 6.48 | 14.72 |
| P4 | 48.29 | 11.58 | 20.42 | 6.95 | 12.75 |
| P5 | 36.33 | 13.59 | 31.33 | 8.38 | 10.37 |

Figure 8 illustrates the valence band (VB) spectra of the pristine and Ar ion irradiated samples. The spectra consist of the density of states (DOS), predominantly influenced by Mo 4d and S 3p states. The peaks indicated by 3 and 4 represent the hybridized states of Mo 4d and S 3p states, and the peak denoted by 1 (situated at 2.7 eV) corresponds to mainly Mo 4d state ($d_z^2$ state) [65]. The VB spectra associated with the occupied Mo 4d-derived states move towards the Fermi level, leading to a displacement of the valence band maximum (VBM) when irradiation fluences increase (Figure 8(b)). These shifts also validated band bending resulting from defect formation. It is noteworthy that at higher irradiation fluences, the characteristics associated with the hybridization of Mo 4d and S 3p-derived states are significantly broadened, suggesting that the alteration of the band structure of $MoS_2$ films leads to the formation of amorphous material at the surface [68]. All peaks in the VB spectra of the P6 sample exhibit broadening, decrease in intensity, and shift away from the Fermi level, indicating substantial degradation of band structure at the highest irradiation fluences.

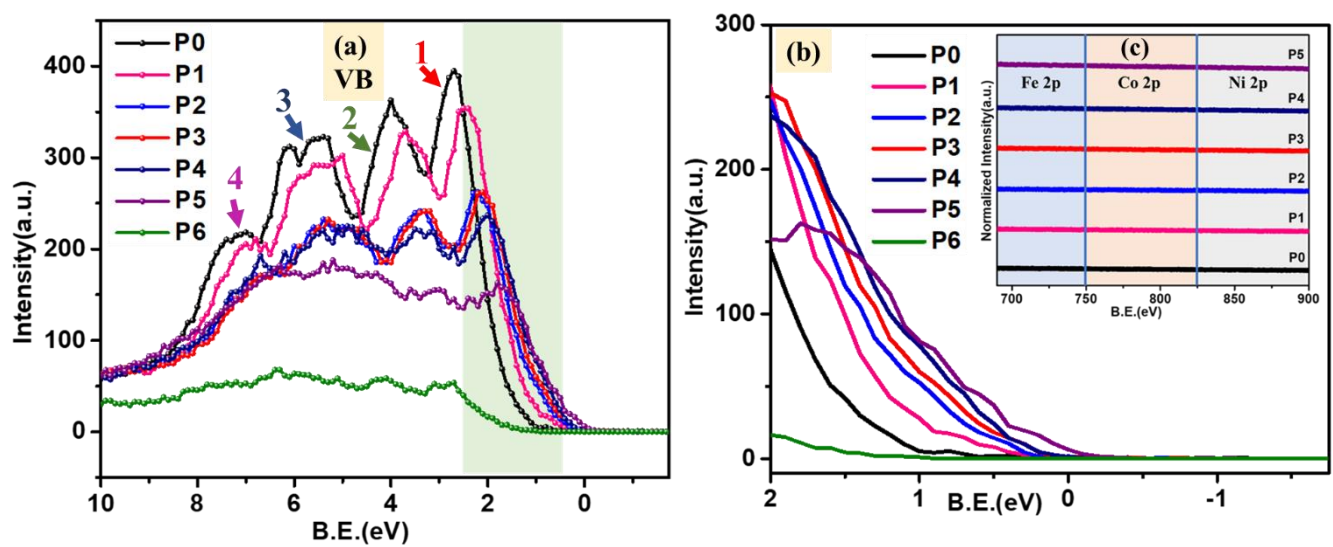

**Figure 8.** (a) Valence band spectra of pristine and $Ar^+$ irradiated samples. (b) A zoom view of the shifting of VB maxima towards the Fermi level. (c) The XPS core level spectra of Fe 2p, Co 2p, and Ni 2p for P0 to P5 samples. It confirms the absence of any magnetic impurities in the samples.

For better understanding of $MoS_2$, we refer the density functional theory (DFT) based theoretical calculations performed in our previous studies [26,32]. It was demonstrated that edge-terminated structure and S-vacancies are the two primary factors to induce ferromagnetism in $MoS_2$. From the detailed calculation, it has been shown that the edge-terminated structure of $MoS_2$ shows ferromagnetism with a magnetic moment is as high as 3.20 $\mu_B$. Further incorporation of S-vacancies in the edge-terminated structure enhances the magnetic moment to 3.85 $\mu_B$, which satisfies the present experimental findings. The structural information is shown in Figure S6 of bulk $MoS_2$, S-vacancy incorporated bulk $MoS_2$, edge-terminated structure, and S-vacancy incorporated edge-terminated structure of $MoS_2$.

### 4. Conclusions:

Nano-dimensional $MoS_2$ thin films have been synthesized on $SiO_2$/Si substrates by the CVD method. The pristine sample exhibits diamagnetic behavior at room temperature, as it contains a comparatively smaller number of edge-terminated structures and S-vacancies. To increase edge-terminated structure and S-vacancies, low-energy Ar ion irradiation has been performed. After ion irradiation, the saturation magnetization value increases gradually with increasing irradiation doses. The formation of S and Mo vacancies is confirmed by Raman spectroscopy data analysis. From FESEM images, it is evident that the irradiated sample with $1\times10^{15}$ ions/cm$^2$ fluence contains a greater number of edge-terminated structures. From the RBS data analysis, it is confirmed that there is a sulfur deficiency in the irradiated samples. XPS study indicates that the S-vacancies and reduced Mo increase at the surface with increasing irradiation fluences. The primary reasons behind the induced magnetism are edge-terminated structure and S-vacancies. Also, as the thickness of the sample is around 20 nm, the defect density is increased than the bulk sample. The sample irradiated with $1\times10^{15}$ ions/cm$^2$ fluence shows the highest saturation magnetization value of 4.18 emu/g. The structural alteration, modifications in electronic structure, and the magnetic properties are in reasonable agreement with each other.

**Acknowledgements:**

The authors thank the ion beam center at Helmholtz Zentrum Dresden Rossendorf (HZDR), Germany for the beam time to irradiate the samples through the "RADIATE" proposal (proposal no.: 22002934). The authors acknowledge the IUAC, New Delhi for the RBS measurements. The authors are grateful to the Physics Department at IIT Delhi for the XRD, Raman, and SQUID facilities; and the Central Research Facilities (CRF) for the FESEM, EPMA, and XPS facilities. Authors also thank Prof. Saswata Bhattacharya for important discussion. One of the authors, Sharmistha Dey, acknowledges IIT Delhi for a Senior Research Fellowship.